# Experimental and Theoretical Study of Thin-covered Composite Dowels considering Multiple Load Conditions

Zhihua Xiong[*,1], Jiaqi Li[1], Xulin Mou[1], Tiankuo Wang[1], Abedulgader Baktheer[2], Markus Feldmann[3]

1. *College of Water Resources and Architectural Engineering, Northwest A&F University, Yangling, China*

2. *Faculty of Civil Engineering and Geodetic Science, Leibniz University Hannover, Hannover, Germany*

3. *Institute of Structural Concrete, RWTH Aachen University, Aachen, Germany*

*\*Corresponding author: zh.xiong@nwsuaf.edu.cn*

**Abstract**：With the widespread application of composite structures in the fields of building and bridge constructions, thin-covered composite dowels are increasingly adopted in various engineering scenarios. This paper presents a design methodology for thin-covered composite dowels, supported by both experimental and theoretical investigations. In the experiment, a novel test rig and specimens are designed to facilitate tensile-shear coupling loading. The study identifies a new failure mode: Restricted Cone Failure (RCF) in thin-covered composite dowels under tensile-shear coupling load, which distinct from conventional composite dowels. This RCF mode is attributed to the thin thickness of the side concrete cover, which restricts the development of the failure cone in the thickness direction. Additionally, a parametric analysis is conducted to evaluate the effects of key factors—such as steel dowel thickness, effective embedment depth, and the tensile strength of steel fiber reinforced concrete—on the bearing capacity and ductility of thin-covered composite dowels. Based on the theoretical findings, comprehensive tensile, shear, and tensile-shear coupling capacity models along with an engineering design model are developed to aid in the practical application of thin-covered composite dowels.

# 1. Introduction

Composite structures have become increasingly popular in the design of buildings and bridges due to their enhanced performance and durability [1-4], in which many new types of connectors have been proposed and studied recently [5-7]. As a type of shear connector in composite structures, composite dowels have gained widespread use for their excellent bearing capacity and ductility, while concrete significantly influences the bearing capacity and failure modes of composite dowels [8-12]. In existing design methods for composite dowels, only the thickness of the top and bottom concrete covers is considered as influencing factors, while the thickness of the side concrete cover is typically regarded as adequate within a specified range [13]. However, with the trend toward lighter steel-concrete composite structures, concrete slab thicknesses particularly in building structures have been continually reduced. This has led to situations where the side concrete cover of composite dowels is thinner, as shown in Fig. 1. For this study, such composite dowels are referred to as *thin-covered composite dowels*. In Fig. 1, conventional composite dowels feature sufficient side concrete cover in Verbund-Fertigteil-Trager (VFT) girders, while thin-covered composite dowels exhibit thin side cover in Modified-Verbund-Fertigteil-Trager (MVFT) girders and steel-concrete composite floors [14-15]. Consequently, it is essential to consider the thickness of the side concrete cover and develop appropriate design methods for thin-covered composite dowels.

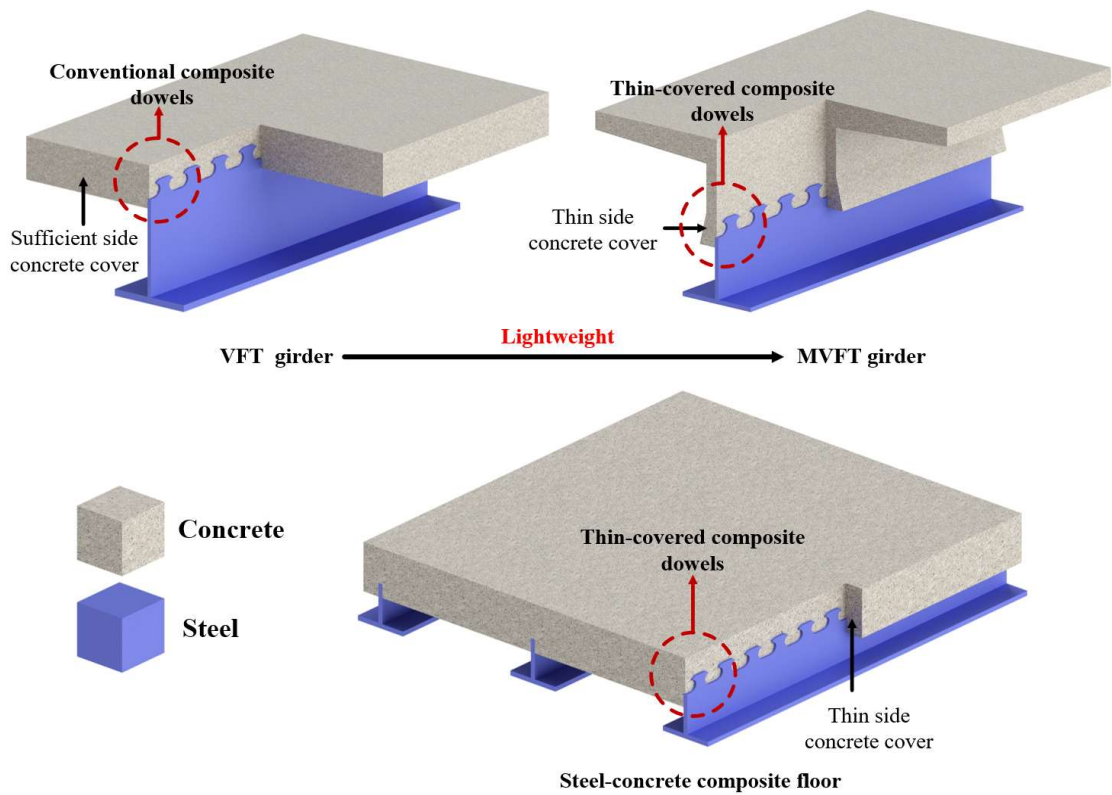

Fig. 1 Application scenes of thin-covered composite dowels

At present, considerable research has been conducted on the bearing capacity and ductility of conventional composite dowels, the shear performance of these dowels remains crucial, as it directly relates to their failure modes. Studies have classified the failure modes of conventional composite dowels into concrete failure [16-18] and steel failure [19-21]. Further categorization of these failures has led to the identification of three primary modes: steel failure, concrete shearing, and prying-out of concrete, with corresponding bearing capacity models proposed [22]. Additionally, composite dowels, especially near the negative bending moment zone of bridges, are often subjected to tensile loads [23]. Numerous tests have shown that concrete failure predominates in these dowels under tensile loads [24-26]. The tensile capacity model for conventional composite dowels experiencing concrete break-out has been established [27-28], and the tensile performance of steel-UHPC composite dowels has also been studied [29]. Furthermore, the performance of composite dowels under tensile-shear coupling loads is critical, with models for tensile-shear coupling capacity developed based on existing tensile and shear capacity models [30-32]. In terms of ductility, a multilinear model has been adopted to describe the relationship between shear force and slip for conventional composite dowels [33]. However, these studies overlook the influence of the side concrete cover thickness.

Despite existing research, there is limited exploration of thin-covered composite dowels. Some experimental and numerical studies have assessed the shear performance of these dowels, revealing that their shear capacity and deformation abilities are reduced compared to conventional dowels, often leading to side concrete cover peeling under shear loading [34-35]. However, these studies focus primarily on shear behavior and omit key factors such as embedment depth.

In general, for conventional composite dowels, the typical failure modes under shear and tensile loads involve cone-shaped concrete pry-out and break-out. In contrast, for thin-covered composite dowels the thin side concrete cover causes a different failure mechanism, such as side concrete cover peeling under shear load [34], which results in lower bearing capacity and ductility. Therefore, this study aims to address the failure modes and bearing capacity of thin-covered composite dowels.

This work presents both experimental and theoretical investigations into the performance of thin-covered composite dowels. The failure mode of these dowels under tensile-shear coupling load is first analyzed through tensile-shear coupling loading tests. A parameter analysis is conducted via numerical simulations to examine the effects of concrete thickness, steel dowel thickness, effective embedment depth, and the tensile strength of Steel Fiber Reinforced Concrete (SFRC) on the bearing capacity and ductility of thin-covered composite dowels. Finally, theoretical analysis is used to discuss the failure modes, and tensile, shear, and tensile-shear coupling capacity models are developed for engineering applications.

## 2. Experimental investigation

### 2.1. Design of specimens

Special specimens with an inclined dowel and a custom test rig were designed to achieve tensile-shear coupling loading of thin-covered composite dowels. In the experiment, four groups in total of eight specimens were designed and fabricated, consisting of Q345 steel plates with thicknesses of 10mm and 12mm, and concrete cube made from C40 and Steel Fiber Reinforced Concrete (SFRC) with thickness of 160mm. The steel dowel spacing $e_x$ was set as 160mm, and the plane size of the concrete cube was 300mm×280mm. The parameters of each specimen are listed in Table 1. The rebars in the specimens were HRB400, consisting of transverse rebars with a diameter of 10mm and stirrups with a diameter of 8mm. The steel dowel was rotated 60° relative to the loading direction to achieve tensile-shear coupling loading, with the shear load of approximately 1.73 times the tensile load. Previous study had shown that the shear capacity of composite dowels is greater than the tensile capacity [31], so this design enabled the composite dowel to reach its ultimate bearing capacity in both the tensile and shear directions as much as possible, resulting in a typical tensile-shear coupling failure mode. The geometric parameters and 3D view of the specimens are shown in Fig. 2, and the material parameters of the specimens are shown in Table 2.

Table 1 test specimens

| Group | $t_w$ | Concrete | Steel |
|---|---|---|---|
| S2-10-C40 | 10 | C40 | Q345 |
| S4-12-C40 | 12 | C40 | Q345 |
| S14-10-SFRC | 10 | SFRC | Q345 |
| S16-12-SFRC | 12 | SFRC | Q345 |

Table 2 Material parameters

| Materials | $f_{cu}$/MPa | $f_y$/MPa | $f_u$/MPa |
|---|---|---|---|
| C40 | 40.3 | —— | —— |
| SFRC | 57.2 | —— | —— |
| Steel | —— | 345 | 550 |

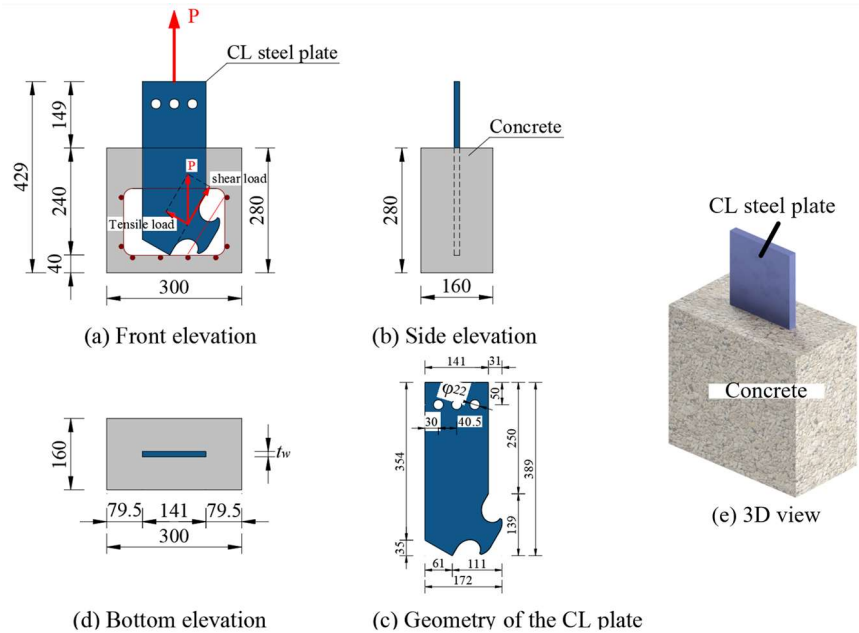

Fig. 2 Geometric parameters of the test specimens

## 2.2. Test rig

As shown in Fig. 3 (a) and (b), the loading device used in the experiment consisted of two clamps, the lower clamp used to fix the specimen and the upper clamp used to apply the load. In order to achieve the loading of the specimen, a special fixture suitable for thin-covered composite dowels was designed, as shown in Fig. 3 (c). The fixture consisted of upper and lower steel plates and four screws that fix them. The upper steel plate was cut into a rectangular area for the steel plate of the specimen to pass through and connect with the upper clamp, and the lower steel plate was drilled for high-strength bolt to pass through and connect with the lower clamp. Before formal loading, the specimen was loaded with a pre-tension force of 5kN, gradually loaded until failure, and then unloaded. During loading, the load and displacement in the experiment were recorded by the loading device, and the crack development of the specimen was manually recorded.

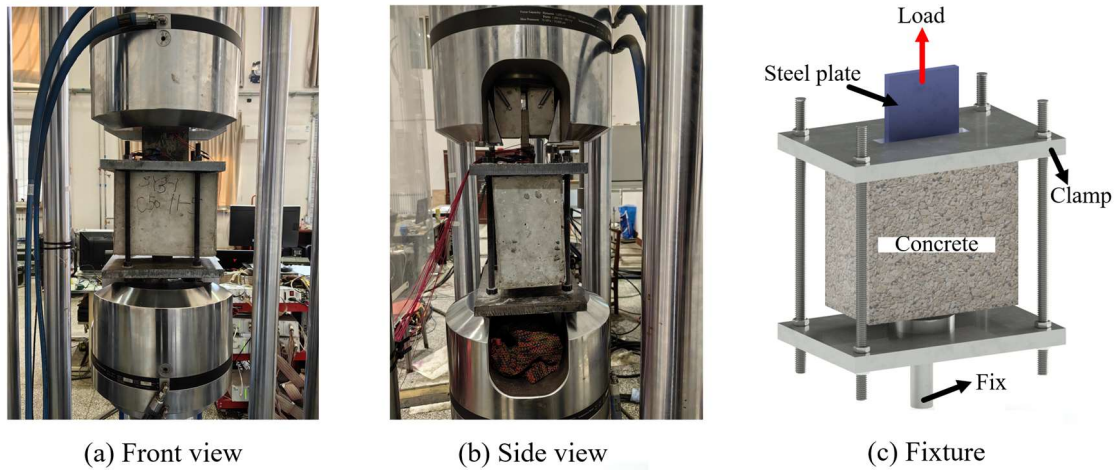

(a) Front view    (b) Side view    (c) Fixture

Fig. 3 Test rig

**2.3. Result and analysis**

*2.3.1.  Concrete crack and failure mode*

The schematic diagram of each view of the specimens is shown in Fig. 4. During the tests of specimens S2-10-C40 and S4-12-C40, as the load increased, cracks first appeared at the top of the front and back views of the specimens, propagating diagonally downward to form the main crack. During this propagation, numerous secondary cracks in various directions also developed. Subsequently, cracks appeared at the top of the left and right views of the specimens and continued to expand downward, forming the main crack (no cracks appeared on the left view of specimen S2-10-C40). Once the main crack on the right view extended sufficiently downward, it began to propagate horizontally to both sides until it reached the edge of the concrete. Simultaneously, the diagonal main cracks on the front and back views gradually extended to the edge of the concrete. Ultimately, the specimens were damaged, and loading was stopped. It was observed that the concrete on the right view of the specimens had partially peeled off. The differences between the specimens resulted in variations in the inclination angles of the main cracks on the front and back views. The cracks in specimens S2-10-C40 and S4-12-C40 are shown in Fig. 5, marked with red lines for clarity.

In the tests of specimens S12-10-SFRC and S14-12-SFRC, as the load increased, cracks appeared at the top of the front and back views of the specimens, extending diagonally downward (no cracks appeared on the front view of specimen S12-10-SFRC). Simultaneously, cracks formed at the top of the right view of the specimens, extending downward and then to both sides after some time. Small cracks also appeared on the top of the left view, though no significant expansion was observed. After stopping the loading, it was observed that some cracks on the front, back, and right views extended to the edge of the concrete, but there was no concrete detachment from the specimens. The cracks in specimens S12-10-SFRC and S14-12-SFRC are shown in Fig. 5, marked with red lines for clarity.

By observing the crack propagation paths of the specimens, it is evident that their failure modes are similar. Specimens S2-10-C40 and S4-12-C40 use C40 concrete, while specimens S12-10-SFRC and S14-12-SFRC use SFRC, which has greater tensile strength and ductility. This difference leads to a more significant failure mode in the C40 specimens. As shown in Fig. 6, the experimental results indicate that the failure mode of thin-covered composite dowels under tensile-shear coupling load is concrete failure. This includes concrete edge failure on the front and back views and concrete pry-out failure on the right view, which is referred to as Restricted Cone Failure (RCF). For thin-covered composite dowels, the thin concrete cover on both sides of the steel dowel restricts the development of the concrete failure cone in the thickness direction, causing cracks to propagate to the front and back views, resulting in RCF as shown in Fig. 6. This failure mode does not occur in conventional composite dowels.

The failure mode of specimens S2-10-C40 and S4-12-C40 is basically consistent with RCF, and the failure area of the concrete is marked with yellow solid lines according to the crack paths of the specimens in Fig. 5. In contrast, due to the higher tensile strength and ductility of SFRC, the crack propagation in specimens S12-10-SFRC and S14-12-SFRC is less extensive. By analyzing the crack paths of these specimens in relation to the RCF model, the failure area of the concrete is indicated with yellow dashed line in Fig. 5.

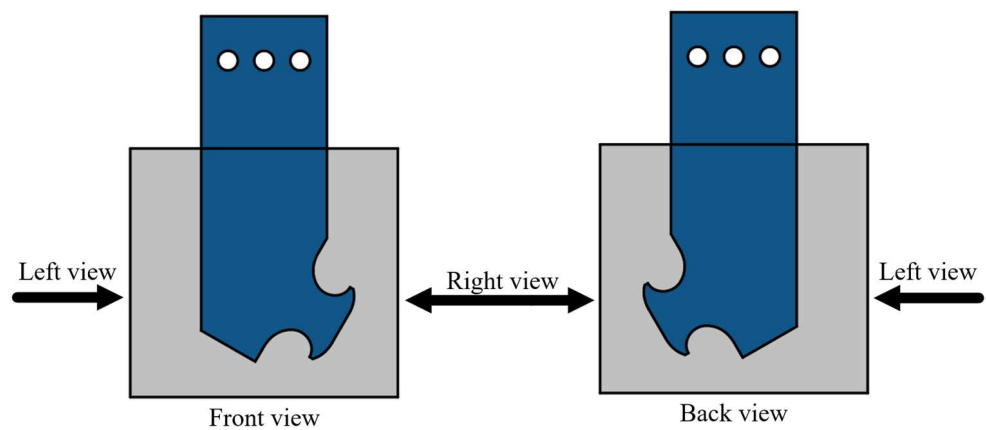

Fig. 4 View sketch map

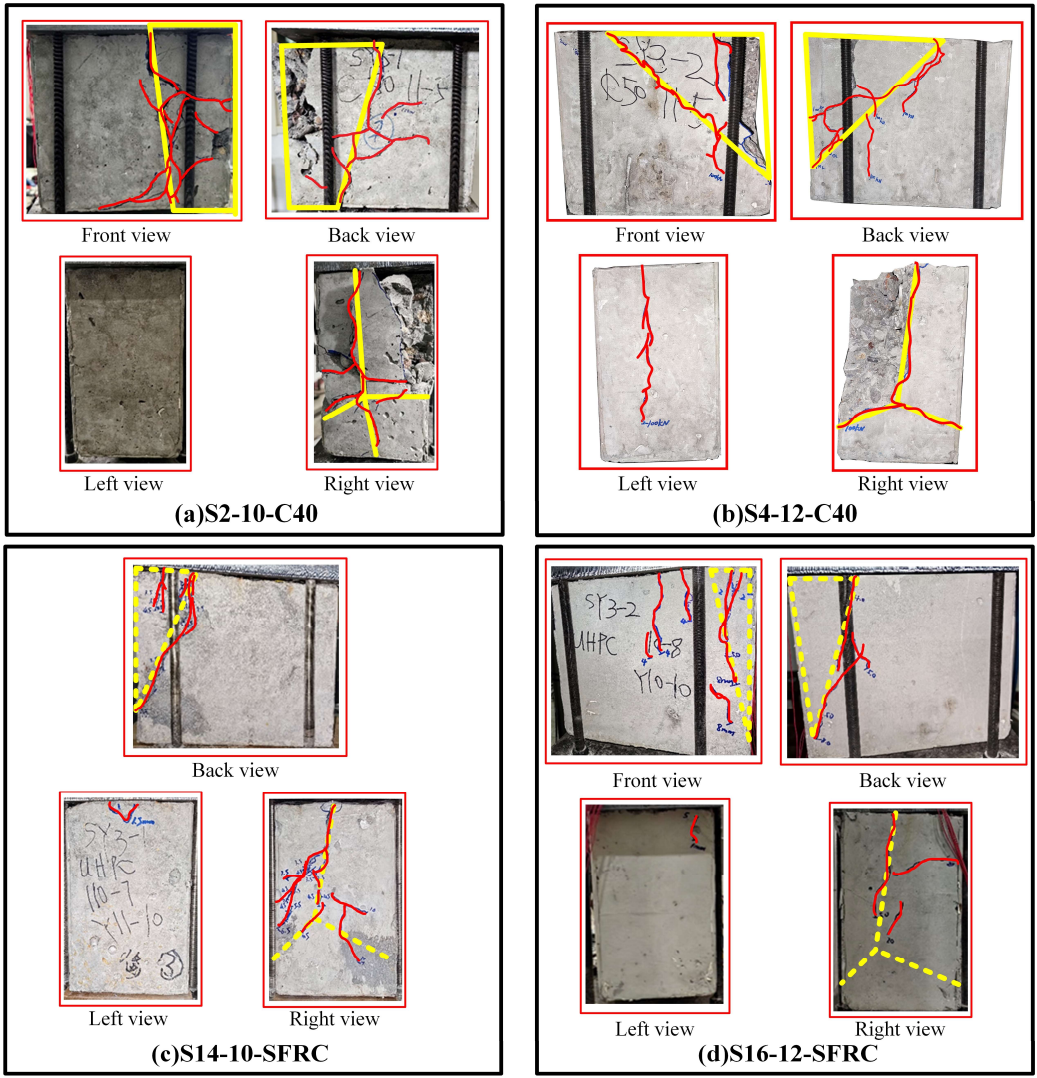

Fig. 5 Crack propagation paths of the specimens

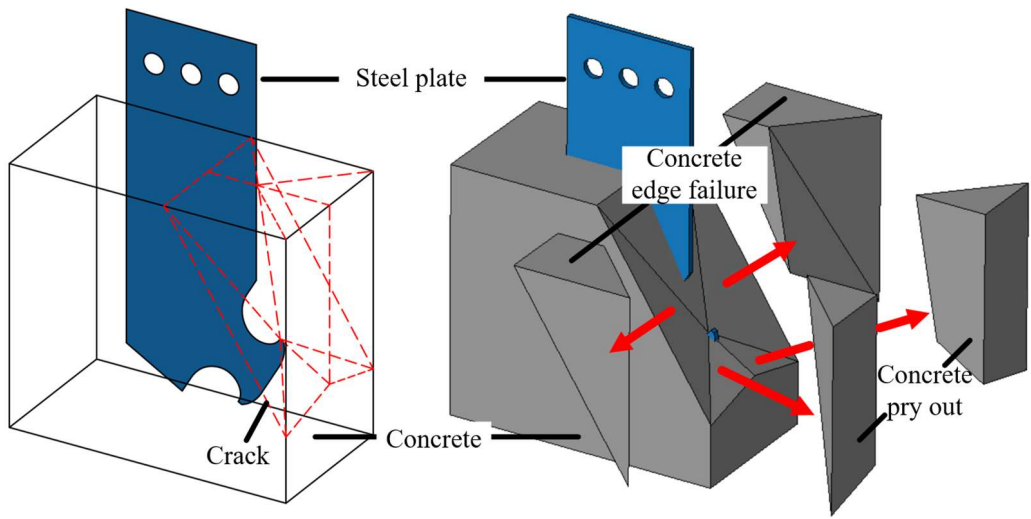

Fig. 6 Restricted Cone Failure mode of thin-covered composite dowels

*2.3.2. Load-displacement behavior*

The load-displacement curves of all specimens are shown in Fig. 7. As illustrated in Fig. 7, increasing the steel dowel thickness enhances the bearing capacity of thin-covered composite dowels. Previous studies have demonstrated that for concrete failure the bearing capacity of conventional composite dowels is independent of the steel dowel thickness [22]. This observation highlights a key difference between the failure modes of thin-covered composite dowels and conventional composite dowels, exhibiting the Restricted Cone Failure (RCF) mechanism described earlier.

Furthermore, improving the performance of the concrete can increase the bearing capacity of conventional composite dowels [37]. To explore the strengthening effect on thin-covered composite dowels, SFRC was used as a substitute for C40 in the experiments. As shown in Fig. 6, the use of SFRC not only significantly improved the bearing capacity of the composite dowels but also enhanced their ductility. The ductility of a composite dowel is defined as the ratio of the displacement at which the bearing capacity decreases to 90% of its peak value to the displacement at the peak bearing capacity. A higher ratio indicates greater ductility. The bearing capacity and ductility values for the specimens are summarized in Table 3.

Table 3 Bearing capacity and ductility of the specimens

| Group | $P_{max}$/kN | $\delta(P_{max})$/mm | $0.9P_{max}$/kN | $\delta(0.9P_{max})$/mm | $\delta(0.9P_{max})/\delta(P_{max})$ |
|---|---|---|---|---|---|
| S2-10-C40 | 75.4 | 3.49 | 67.86 | —— | —— |
| S4-12-C40 | 108.7 | 2.29 | 97.83 | 2.76 | 1.205 |
| S12-10-SFRC | 100.8 | 4.50 | 90.72 | 5.78 | 1.284 |
| S14-12-SFRC | 140.6 | 5.53 | 126.54 | 8.93 | 1.615 |

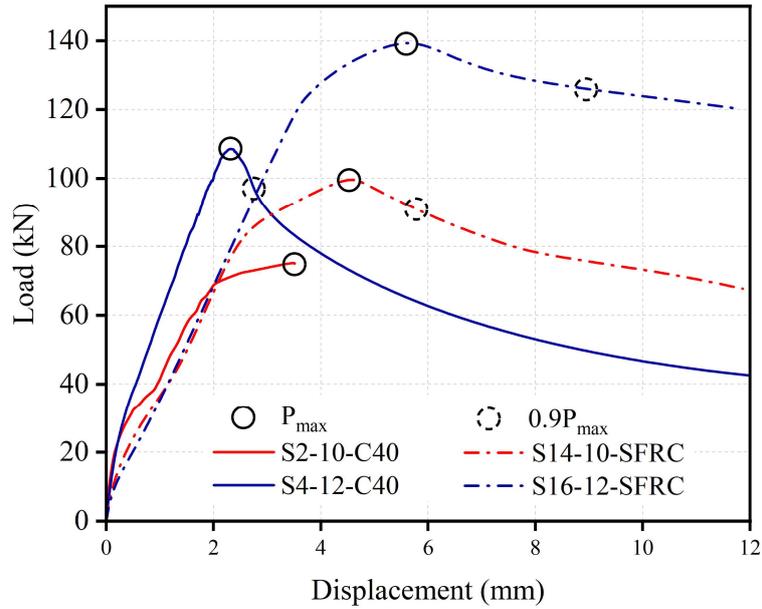

Fig. 7 Load-slip curve of specimens

## 3. Parameter analysis

### 3.1. Parameter design

Numerical simulations are conducted to perform a parametric analysis of the pure tensile, pure shear, and tensile-shear coupling capacities of thin-covered composite dowels. Two types of steel dowels, CL (clothoid) and PZ (puzzle) shapes were selected, and their geometrical parameters are illustrated in Fig. 8. As shown in Fig. 9, the geometrical parameters of the thin-covered composite dowels are as follows: the dowel spacing $e_x$ is 160 mm, and the plane size of the concrete slab is 400 mm × 400 mm. The HRB400 rebars used in the dowels have a diameter of 10 mm for both the transverse rebars and the stirrups.

Based on the previous experiments, it can be concluded that, for thin-covered composite dowels the bearing capacity is influenced by the steel dowel thickness $t_w$, which can be effectively enhanced using SFRC. Moreover, according to the RCF mechanism discussed earlier, the concrete thickness $b_c$ and the top concrete cover $c_t$ are expected to influence the failure range of the concrete, thereby affecting the bearing capacity of thin-covered composite dowels.

Additionally, various loading angles $\theta$ were selected to simulate pure tensile, pure shear, and tensile-shear coupling loading conditions, where $\theta=0°$ represents pure shear, and $\theta=90°$ represents pure tension. Therefore, the influence of several parameters—including steel dowel thickness $t_w$, concrete thickness $b_c$, top concrete cover $c_t$, steel dowel type, concrete type, and loading angle $\theta$—on the bearing capacity of thin-covered composite dowels was investigated. The parameter scheme is detailed in Table 4, in which is comprised 296 structural configurations.

Table 4 Parameter scheme

| Type of steel dowel | Concrete | $t_w$(mm) | $b_c$(mm) | $c_t$(mm) | $\theta°$ |
|---|---|---|---|---|---|
|  |  | 4-20 | 100-240 | 10-70 | 0° |
|  | NC(C50) | 4-20 | 100-240 | 10-70 | 90° |
| CL type/PZ type |  | 6-20 | 100-240 | 20-70 | 0° -90° |
|  | SFRC | 12 | 100 | 70 | 0° |
|  |  | 12 | 100 | 70 | 90° |
| Total |  |  | 296 |  |  |

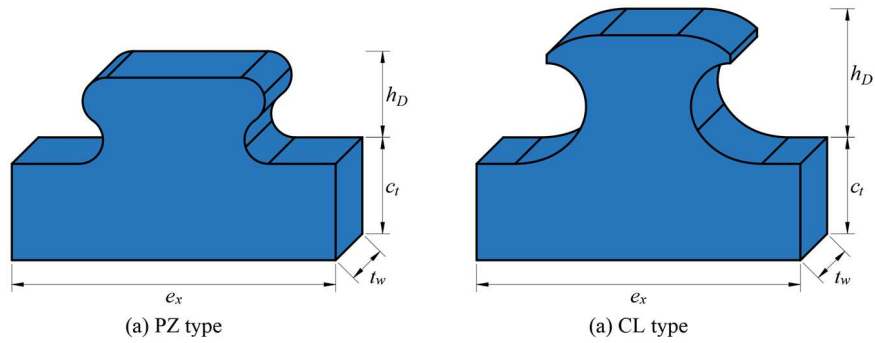

(a) PZ type  (a) CL type

Fig. 8 Two types of steel dowel

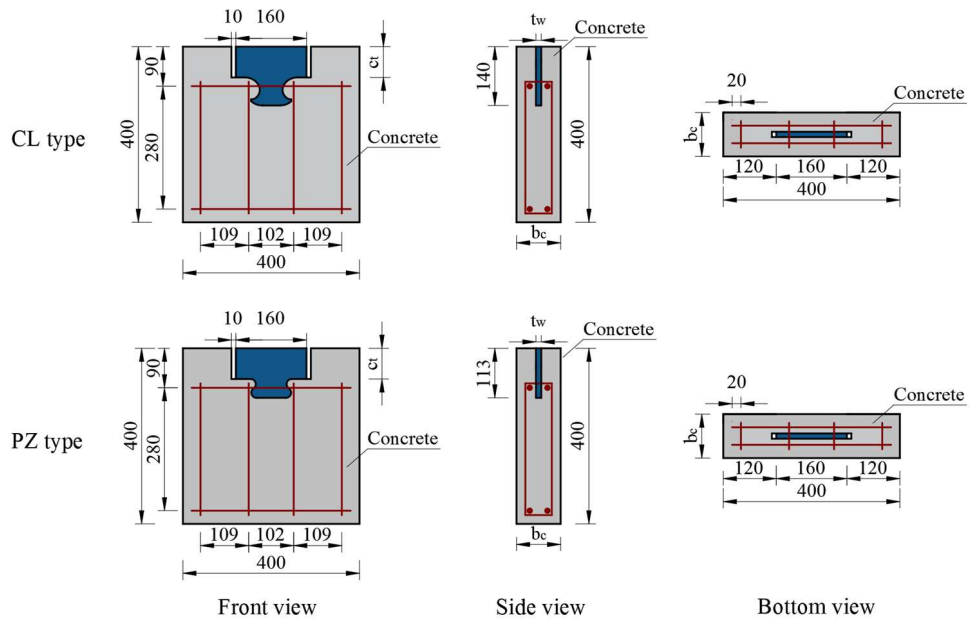

Fig. 9 Geometric parameter of thin-covered composite dowels

## 3.2. Model validation

The experimental results are compared with those from numerical simulations to verify the accuracy of the finite element model for thin-covered composite dowels. As shown in Fig. 10, the finite element model comprises three main components: a CL-shaped steel dowel, concrete and rebars. The concrete was modeled using the concrete damage plasticity model recommended in GB50010-2010 [36], while the steel was modeled using a bilinear hardening model. The material properties for each component were consistent with those used in the experiments.

The solid element C3D8R was used to simulate the concrete and steel plate, and the truss element T3D2 was used to simulate the rebars, with a mesh size of 10mm for both. The contact between steel plate and concrete was set as surface-to-surface contact, and the contact characteristics are simulated using normal hard contact and tangential penalty function, with a friction coefficient of 0.3. The rebars were embedded within the surrounding concrete using the embedded region constraint. The degrees of freedom of the top surface of the concrete were fully constrained, while the top surface of the steel plate was coupled to a reference point for loading. A vertical upward displacement loading mode was applied. In the FE model, the "Static, General" module was adopted, and concrete tensile damage, *i.e.* DAMAGET, was added to the field output to display the crack development of the concrete, which depends on the tensile damage-strain relationship set in its constitutive model. The concrete tensile damage varies from 0 to 1, where 0 represents no damage and 1 represents complete damage.

As illustrated in Fig. 11, the load-displacement curves of specimen S4-12-C40 obtained from the experiment and numerical simulation demonstrate close agreement, with a difference in bearing capacity of only 1.2%. Furthermore, as shown in Fig. 12, the failure mode of the specimen in the numerical simulation closely matches that observed in the experiment. These results validate the accuracy and reliability of the numerical simulation in studying the bearing capacity of thin-covered composite dowels.

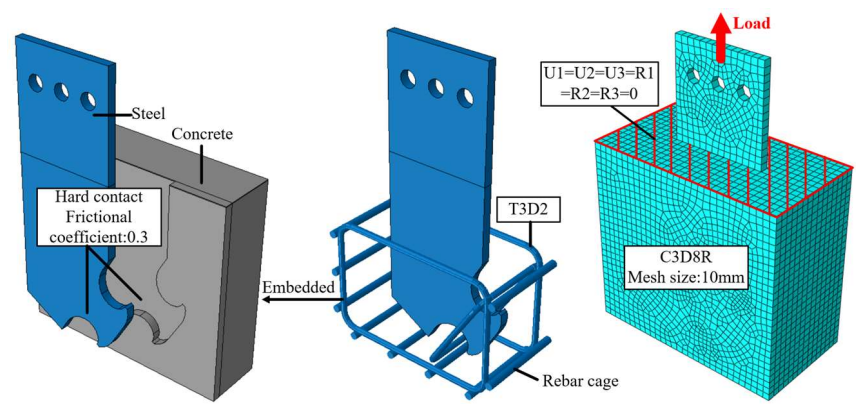

Fig. 10 Finite element model of test specimen.

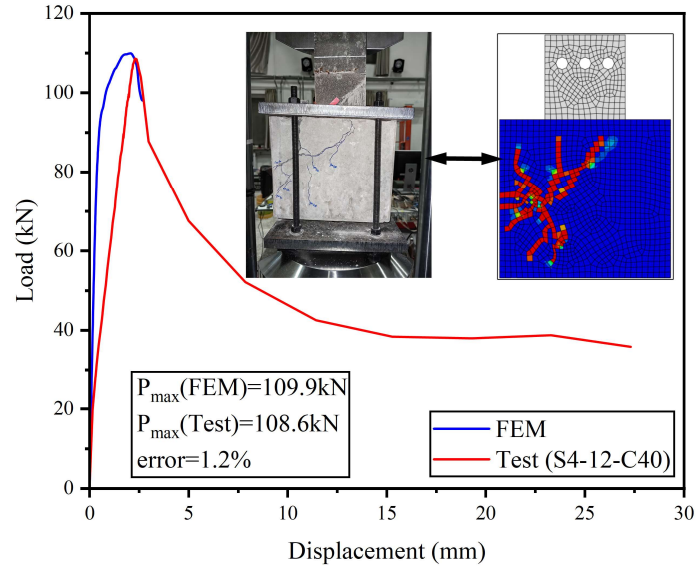

Fig. 11 Comparison of load-displacement curves between experiment and FEM

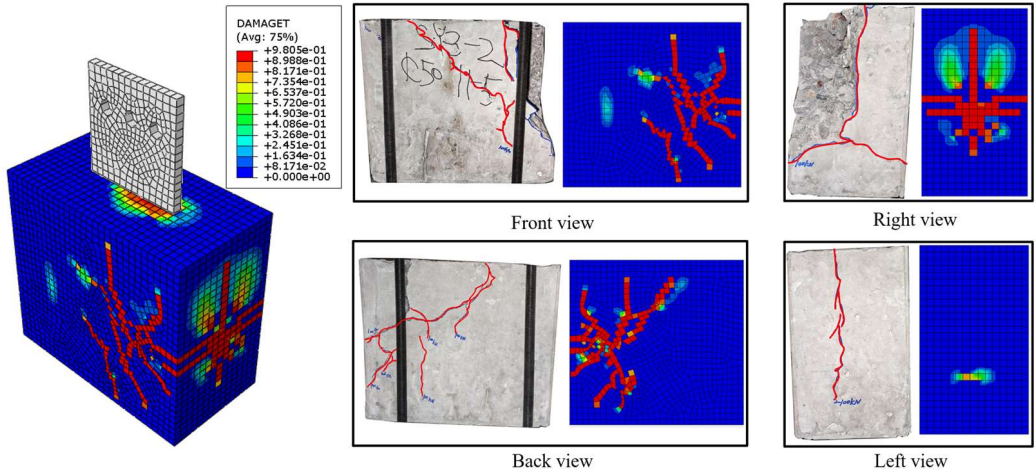

Fig. 12 Comparison of failure mode between experiment and FEM

## 4. Pure tensile capacity of thin-covered composite dowels

### 4.1. Influence of steel dowel thickness $t_w$ and concrete thickness $b_c$

To investigate the effects of concrete thickness ($b_c$) and steel dowel thickness ($t_w$) on the tensile capacity of thin-covered composite dowels, parametric analyses were conducted. The analyses considered nine concrete thicknesses within the range of 100–200 mm and seven steel dowel thicknesses within the range of 4–20 mm, covering the typical size range for engineering applications, as shown in Table 5. All other parameters were held constant: the top concrete cover

($c_t$) was 70 mm, and the concrete used was C50, with an elastic modulus of 34,500 MPa, Poisson's ratio of 0.2, uniaxial compressive strength of 32.4 MPa, and uniaxial tensile strength of 2.6 MPa. The constitutive model of the concrete was consistent with that described in Section 3.2.

Figure 13(a) illustrates the tensile capacity of CL and PZ thin-covered composite dowels under varying $b_c$. It can be observed that the tensile capacity increases with $b_c$ (100–240 mm), showing a 42% increase for CL composite dowels and a 40% increase for PZ composite dowels. As shown in Fig. 13(b), composite dowels of the same shape exhibit nearly identical initial stiffness, which remains unaffected by $b_c$.

Figure 14(a) shows the tensile capacity of CL and PZ thin-covered composite dowels under varying $t_w$. The tensile capacity increases with $t_w$ (4–20 mm), with a 62% increase observed for CL composite dowels and a 32% increase for PZ composite dowels. Figure 14(b) reveals that the initial stiffness of thin-covered composite dowels increases with $t_w$, while their ductility decreases as $t_w$ increases. As shown in Fig. 15, the tensile capacity of thin-covered composite dowels exhibits an exponential relationship with $b_c$ and $t_w$, expressed as follows:

$$F_{T,CL,PZ} \sim b_c^{0.363} \qquad (1)$$

$$F_{T,CL} \sim t_w^{0.280} \qquad (2)$$

$$F_{T,PZ} \sim t_w^{0.185} \qquad (3)$$

Table 5 Parameter scheme for concrete thickness $b_c$ and steel dowel thickness $t_w$

| Parameter | Size/mm |
|---|---|
| $b_c$ | 100, 120, 140, 160, 180, 190, 200, 220, 240 |
| $t_w$ | 4, 5, 6, 8, 10, 12, 20 |

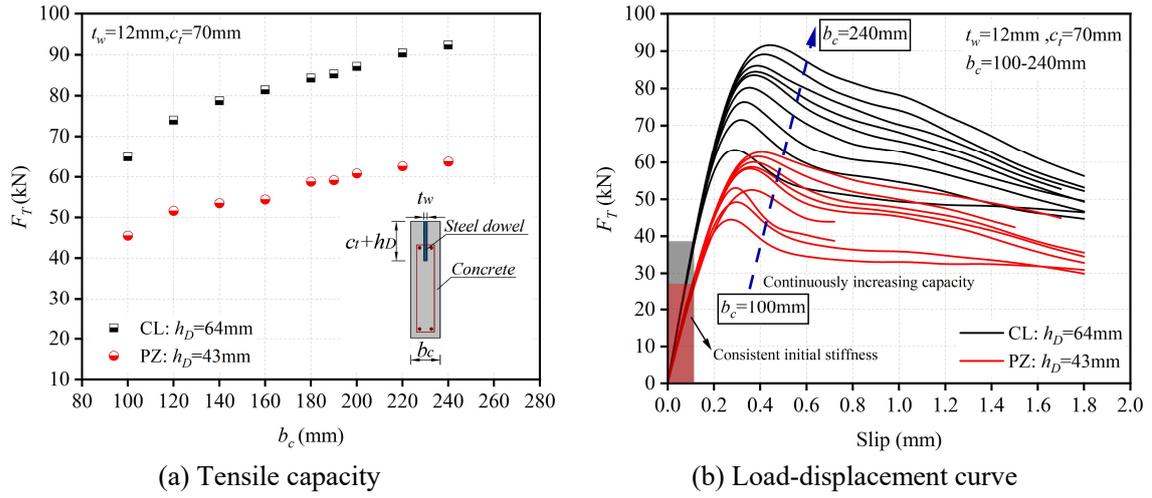

(a) Tensile capacity  (b) Load-displacement curve

Fig. 13 Comparison of tensile capacity with different concrete thicknesses

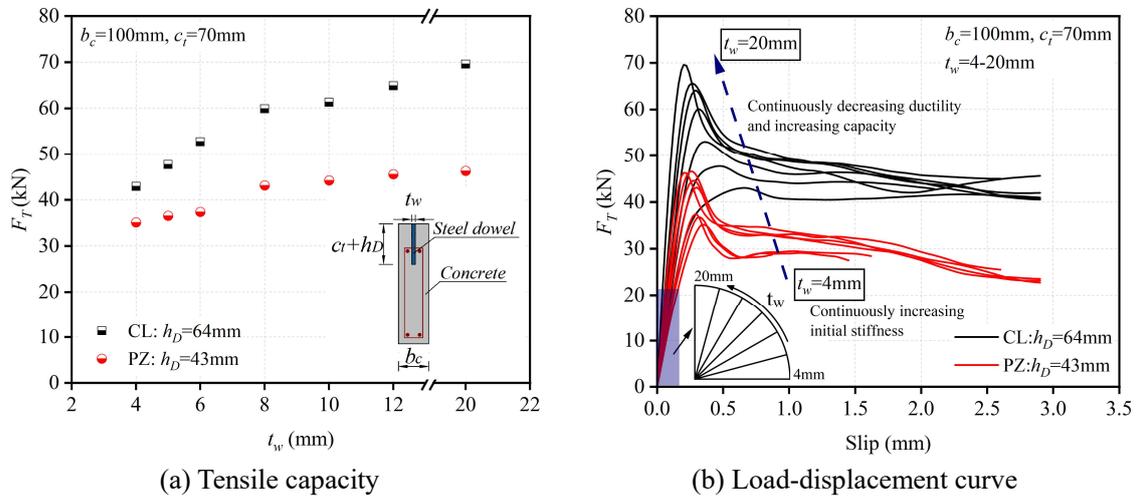

(a) Tensile capacity  (b) Load-displacement curve

Fig. 14 Comparison of tensile capacity with different steel thicknesses

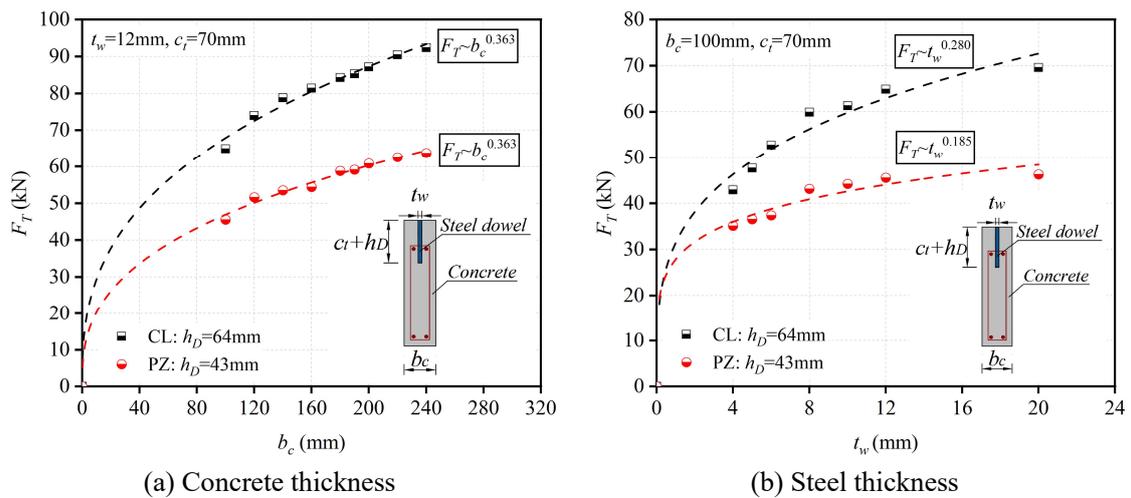

(a) Concrete thickness  (b) Steel thickness

Fig. 15 Influence of steel dowel thickness and concrete thickness on the tensile capacity

## 4.2. Influence of the effective embedment depth $h_{ef}$

For conventional concrete composite dowels, the tensile capacity is positively correlated with the effective embedment depth of steel dowels [28]. A similar relationship is assumed to apply for thin-covered composite dowels. The relationship between the effective embedment depth ($h_{ef}$) and the top concrete cover ($c_t$) is expressed in Eq. (4).

To investigate the effect of the effective embedment depth $h_{ef}$ on the tensile capacity of thin-covered composite dowels, six top concrete cover values $c_t$ ranging from 10 mm to 70 mm were selected and converted into corresponding effective embedment depths $h_{ef}$, as shown in Table 6. All other parameters were kept constant: the concrete thickness $b_c$ is 100 mm, the steel dowel thickness $t_w$ is 12 mm, and the concrete is C50, as detailed in Table 6. As illustrated in Fig. 16, the tensile capacity of thin-covered composite dowels increases exponentially with $h_{ef}$. The relationships are expressed in Eq. (5) for CL composite dowels and Eq. (6) for PZ composite dowels.

Moreover, the tensile capacity of the composite dowels is influenced by the effective distance between the load points of the steel dowels. This effective distance is defined as the distance between the two ends of the steel dowels at the effective embedment depth. The equations for calculating the effective distance are given in Eq. (7) for CL dowels and Eq. (8) for PZ dowels.

$$h_{ef} = \begin{cases} c_t + 0.625 \cdot h_D & \text{PZ-dowel} \\ c_t + 0.75 \cdot h_D & \text{CL-dowel} \end{cases} \quad (4)$$

$$F_{T,CL} \sim h_{ef}^{0.55} \quad (5)$$

$$F_{T,PZ} \sim h_{ef}^{0.68} \quad (6)$$

$$L_{e,CL} = 0.464 e_x \quad (7)$$

$$L_{e,PZ} = 0.554 e_x \quad (8)$$

Table 6 Parameter scheme for effective embedment depths $h_{ef}$

| type | $c_t$(mm) | $h_D$(mm) | $h_{ef}$(mm) |
|---|---|---|---|
| CL | 10 | 64 | 58 |
| | 20 | | 68 |
| | 30 | | 78 |
| | 40 | | 88 |
| | 50 | | 98 |
| | 70 | | 118 |
| PZ | 10 | 43 | 36.875 |
| | 20 | | 46.875 |
| | 30 | | 56.875 |
| | 40 | | 66.875 |
| | 50 | | 76.875 |
| | 70 | | 96.875 |

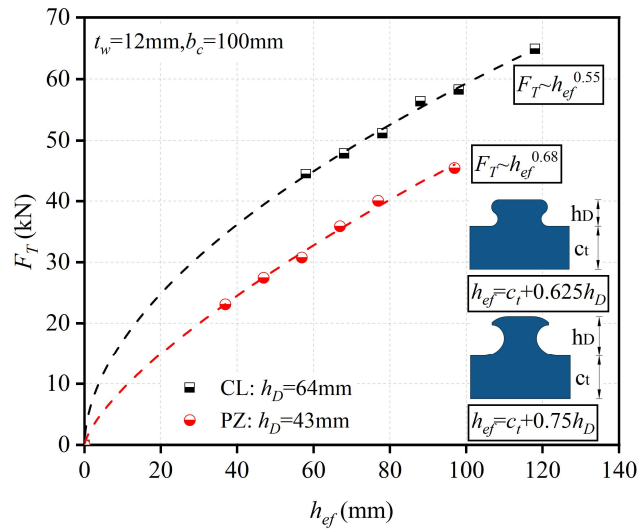

Fig. 16 Influence of effective embedment depth on the tensile capacity

**4.3. Influence of the tensile strength of SFRC**

Steel Fiber Reinforced Concrete (SFRC) is made by adding steel fibers to concrete, which effectively enhances both its tensile strength and ductility [38]. To evaluate the effect of SFRC on the tensile capacity of thin-covered composite dowels, three types of SFRC with different tensile strengths are selected, while keeping other parameters such as compressive strength unchanged. The uniaxial tensile strengths of the SFRC samples are 4 MPa, 5 MPa, and 6 MPa, respectively, with a uniaxial compressive strength of 32.4 MPa. SFRC was simulated using the concrete damaged plasticity model, and its constitutive behavior was adopted from Gowriplan [39].

All other parameters of the composite dowel were held constant: the concrete thickness $b_c$ was 100 mm, the steel dowel thickness $t_w$ was 12 mm, and the top concrete cover $c_t$ was 70 mm. As shown in Fig. 17, the tensile capacity of thin-covered composite dowels increases linearly with the tensile strength of SFRC. The relationships are expressed in Eq. (9) for CL composite dowels and Eq. (10) for PZ composite dowels.

$$F_{T,CL} \sim (0.154 \cdot f_{tk} + 1.558) \quad (9)$$

$$F_{T,PZ} \sim (0.21 \cdot f_{tk} + 1.484) \quad (10)$$

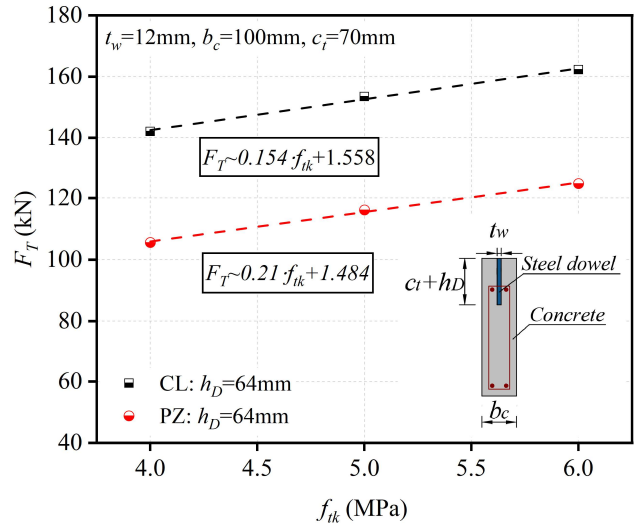

Fig. 17 Influence of tensile strength of SFRC on the tensile capacity

### 4.4. Pure tensile capacity model

For conventional composite dowels, the tensile capacity model shows that under pure tension the failure of concrete usually occurs [27-28], characterized by the formation of a concrete break-out cone on the top surface of the concrete as shown in Fig. 18. In this failure mode, the tensile capacity of conventional composite dowels is directly proportional to the square of the effective embedment depth of the steel dowel, the square root of the compressive strength of the concrete, and the size effect factor as shown in Eq. (11)-(13).

$$F_t \sim h_{ef}^2 \qquad (11)$$

$$F_t \sim \sqrt{f_{ck}} \qquad (12)$$

$$F_t \sim \frac{1}{\sqrt{1+\dfrac{h_{ef}}{100}}} \qquad (13)$$

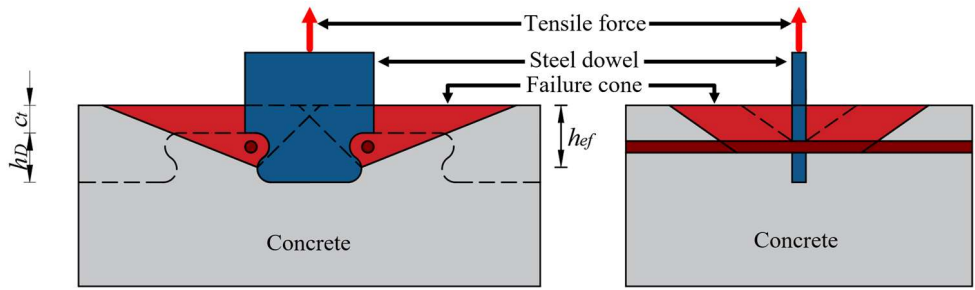

Fig. 18 Model for tensile behavior of conventional composite dowels with concrete break out

So far, there has been no research addressing the pure tensile capacity of thin-covered composite dowels. For conventional composite dowels, the tensile capacity is related to the size of the concrete break-out cone and the concrete strength. However, for thin-covered composite dowels, the thin concrete thickness limits the full expansion of the break-out cone in the thickness direction. As a result, the concrete failure extends to the concrete side surface, as depicted in Fig. 19. Therefore, a new tensile capacity model for thin-covered composite dowels is required.

For the pure tensile capacity model of thin-covered composite dowels, the relationship between tensile capacity and the size of the concrete break-out cone in conventional composite dowels is transformed into a relationship involving the effective embedment depth, effective distance, concrete thickness, and steel dowel thickness, while retaining the dependence on concrete strength. This assumes that the relationship between bearing capacity and concrete strength is consistent across different failure modes (i.e., concrete failure), as demonstrated in previous studies [22,27].

Summarizing Sections 4.1–4.4, the complete tensile capacity models for thin-covered composite dowels with ordinary concrete are given in Eq. (14) and Eq. (15), while those for dowels with SFRC are presented in Eq. (16) and Eq. (17). In the equations, $f_{ck}$ and $f_{tk}$ respectively represent the standard values of axial compressive and tensile strength of concrete. The comparison between the FEA results and model predictions is shown in Fig. 20. The results indicate that the theoretical model predictions align well with the FEA results, demonstrating the model's good predictive accuracy for pure tensile capacity.

$$F_{T,CL} = 1.054 \cdot b_c^{0.363} \cdot t_w^{0.28} \cdot h_{ef}^{0.55} \cdot \sqrt{f_{ck}} \cdot L_{e,CL} \qquad (14)$$

$$F_{T,PZ} = 0.478 \cdot b_c^{0.363} \cdot t_w^{0.185} \cdot h_{ef}^{0.68} \cdot \sqrt{f_{ck}} \cdot L_{e,PZ} \qquad (15)$$

$$F_{T,CL} = 1.054 \cdot (0.15395 \cdot f_{tk} + 1.55836) \cdot b_c^{0.363} \cdot t_w^{0.28} \cdot h_{ef}^{0.55} \cdot \sqrt{f_{ck}} \cdot L_{e,CL} \qquad (16)$$

$$F_{T,PZ} = 0.478 \cdot (0.21004 \cdot f_{tk} + 1.48387) \cdot b_c^{0.363} \cdot t_w^{0.185} \cdot h_{ef}^{0.68} \cdot \sqrt{f_{ck}} \cdot L_{e,PZ} \qquad (17)$$

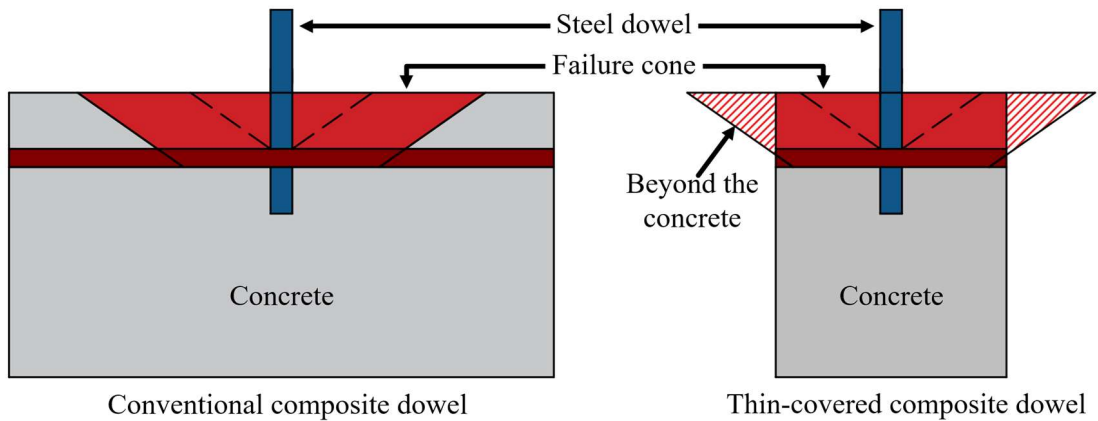

Fig. 19 Limitation of insufficient concrete thickness on the concrete break-out cone

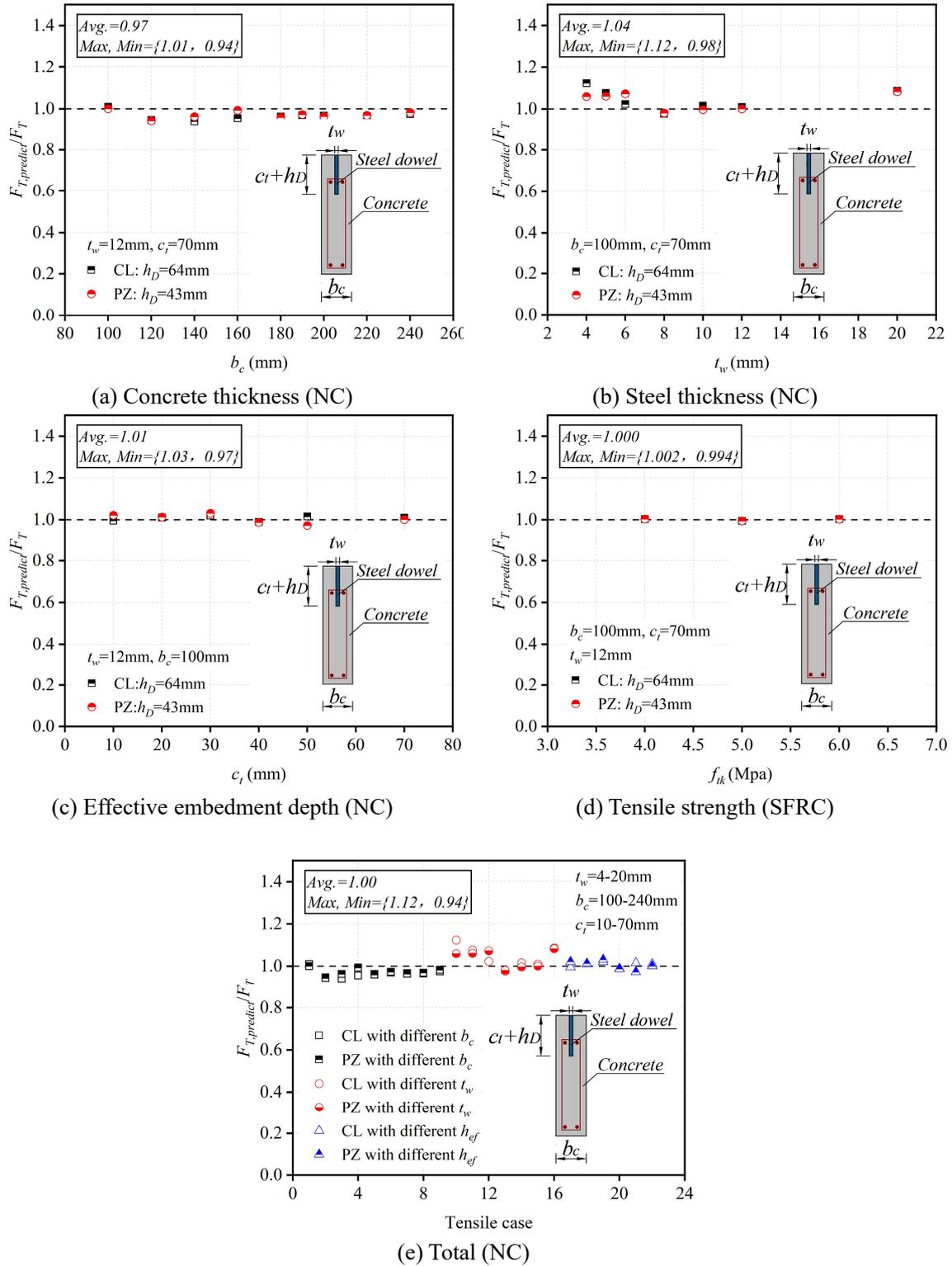

Fig. 20 Comparison between the FEA and the model prediction results

## 5. Pure shear capacity of thin-covered composite dowels

### 5.1. Influence of steel dowel thickness $t_w$ and concrete thickness $b_c$

To investigate the influence of concrete thickness $b_c$ and steel dowel thickness $t_w$ on the shear capacity of thin-covered composite dowels, the same parameter scheme as in Section 4.1 is adopted. The concrete thickness $b_c$ ranges from 100 mm to 240 mm, while the steel dowel thickness $t_w$ varies from 4 mm to 20 mm. The top concrete cover $c_t$ is set to 70 mm, and the concrete is C50.

Fig. 21(a) shows the shear capacity of CL and PZ composite dowels with varying $b_c$. It can be observed that the shear capacity of thin-covered composite dowels increases with $b_c$ (100–240 mm), with a 79% increase for CL composite dowels and 121% for PZ composite dowels. As seen in Fig. 21(b), thin-covered composite dowels of the same shape exhibit essentially consistent initial stiffness that does not change with $b_c$, but the ductility of these dowels ascends as $b_c$ increases.

Fig. 22(a) illustrates the shear capacity of thin-covered composite dowels with different $t_w$. The shear capacity increases with $t_w$ (4–20 mm), with a 101% increase for CL composite dowels and 68% for PZ composite dowels. As shown in Fig. 22(b), the initial stiffness of thin-covered composite dowels increases with $t_w$, but their ductility decreases as $t_w$ increases.

From Fig. 23, it is found that the shear capacity of thin-covered composite dowels follows a certain exponential relationship with $b_c$ and $t_w$, as expressed in the following equations:

$$F_{s,CL} \sim b_c^{0.705} \quad (18)$$

$$F_{s,PZ} \sim b_c^{0.983} \quad (19)$$

$$F_{s,CL} \sim t_w^{0.398} \quad (20)$$

$$F_{s,PZ} \sim t_w^{0.331} \quad (21)$$

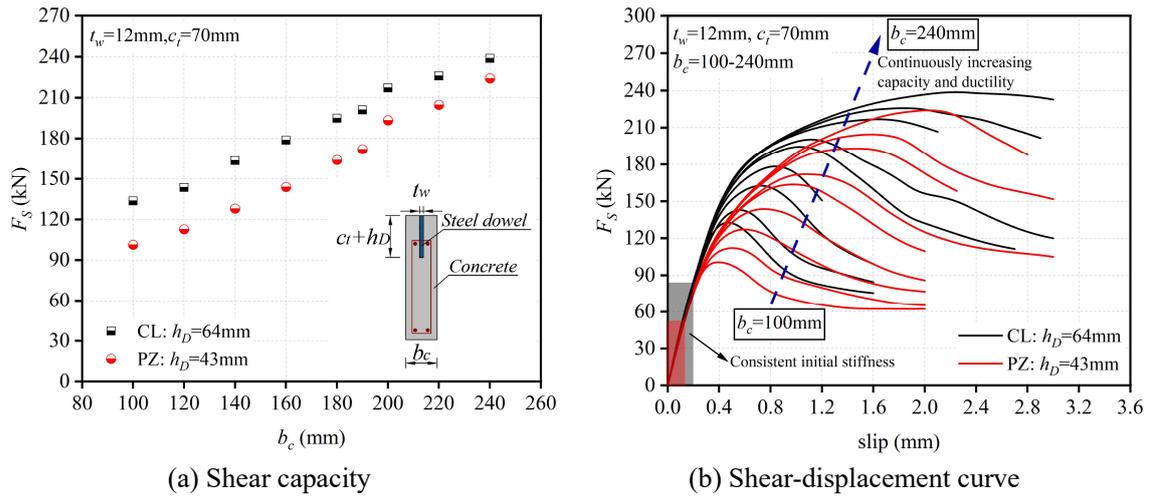

(a) Shear capacity   (b) Shear-displacement curve

Fig. 21 Comparison of shear capacity with different concrete thicknesses

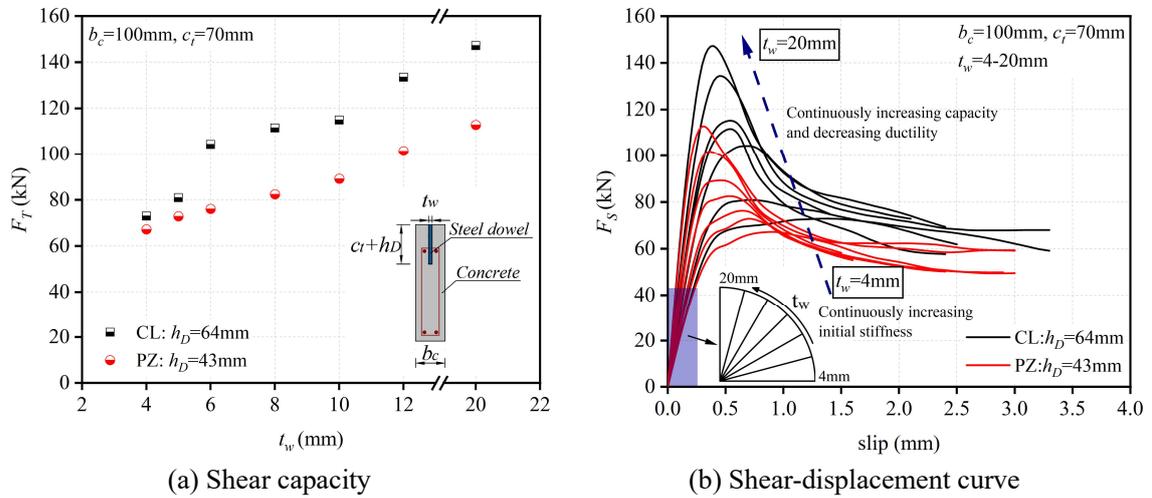

(a) Shear capacity   (b) Shear-displacement curve

Fig. 22 Comparison of shear capacity with different steel thicknesses

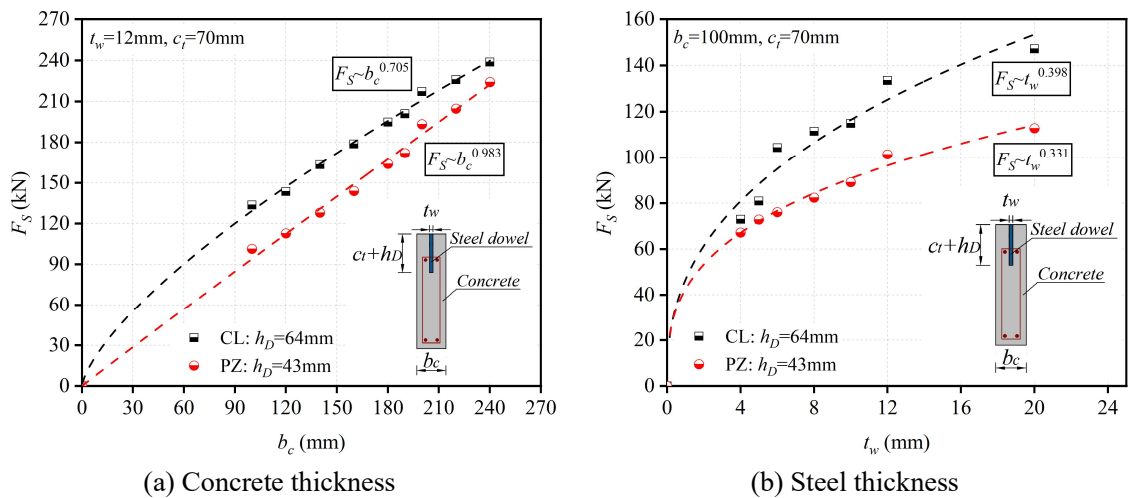

(a) Concrete thickness   (b) Steel thickness

Fig. 23 Influence of steel dowel thickness and concrete thickness on the shear capacity

## 5.2. Influence of the effective embedment depth $h_{ef}$

As discussed in Section 4.2, there is an exponential relationship between the tensile capacity of thin-covered composite dowels and the effective embedment depth of the steel dowel. A similar relationship can be assumed for shear capacity.

To explore the impact of the effective embedment depth on the shear capacity of thin-covered composite dowels, a parameter scheme similar to that in Table 6 is adopted. The range of the top concrete cover $c_t$ is 10–70 mm, the concrete thickness $b_c$ is fixed at 100 mm, the steel dowel thickness $t_w$ is 12 mm, and the concrete is C50, consistent with the scheme in Section 4.2. As shown in Fig. 24, the shear capacity of thin-covered composite dowels increases exponentially with $h_{ef}$. The relationship between shear capacity and effective embedment depth $h_{ef}$ is expressed in Eq. (22) for CL composite dowels and Eq. (23) for PZ composite dowels.

$$F_{S,CL} \sim h_{ef}^{0.23} \quad (22)$$

$$F_{S,PZ} \sim h_{ef}^{0.26} \quad (23)$$

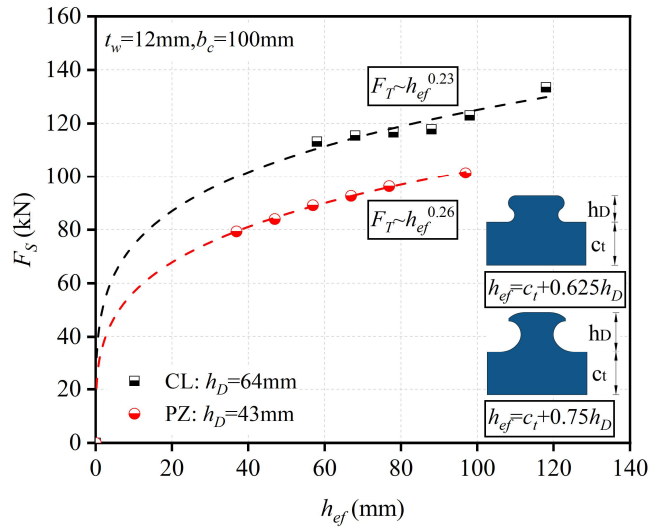

Fig. 24 Influence of effective embedment depth on the shear capacity

## 5.3. Influence of the tensile strength of SFRC

As discussed in Section 4.3, the use of SFRC effectively improves the tensile capacity of thin-covered composite dowels. The same parameter scheme is adopted here to study its effect on shear capacity. The tensile strength of SFRC ranges from 4 MPa to 6 MPa, with the concrete thickness $b_c$ set to 100 mm, steel dowel thickness $t_w$ at 12 mm, and top concrete cover $c_t$ at 70 mm.

Fig. 25 shows that the shear capacity of thin-covered composite dowels increases linearly with the tensile strength of SFRC. This relationship is expressed in Eq. (24) for CL composite dowels and Eq. (25) for PZ composite dowels.

$$F_{S,CL} \sim 0.114 \cdot f_{tk} + 1.596 \qquad (24)$$

$$F_{S,PZ} \sim 0.185 \cdot f_{tk} + 1.705 \qquad (25)$$

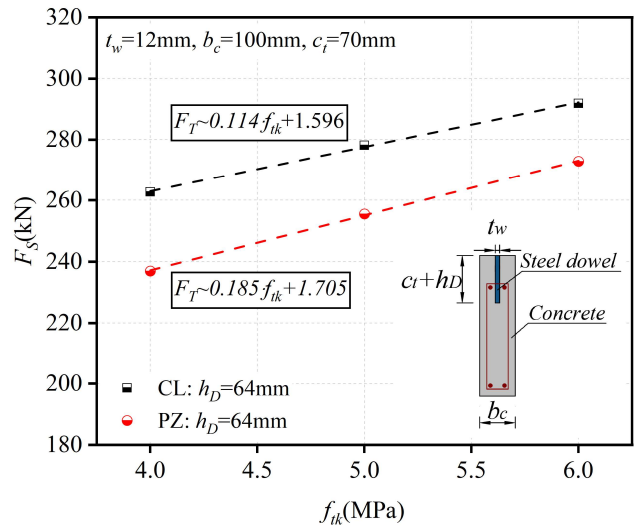

Fig. 25 Influence of tensile strength of SFRC on the shear capacity

### 5.4. Pure shear capacity model

For conventional composite dowels, a shear capacity model has been proposed [22]. As shown in Fig. 26, when the top or bottom concrete cover of a composite dowel is insufficient, the hydrostatic pressure at the load introduction zone generates transverse tension forces, which lead to a conically shaped concrete pry-out, resulting in a concrete pry-out failure mode. In this failure mode, the shear capacity of the composite dowel is directly proportional to the 1.5 power of the height of the concrete pry-out cone and the square root of the concrete compressive strength. It is independent of the concrete thickness and steel dowel thickness, as expressed in Eq. (26) and Eq. (27).

$$F_S \sim h_{po}^{1.5} \quad (26)$$

$$F_S \sim \sqrt{f_{ck}} \quad (27)$$

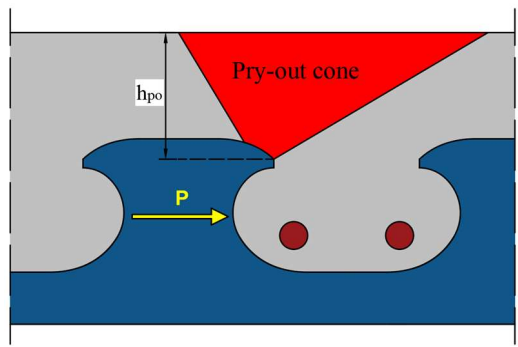

Fig. 26 Concrete pry-out of conventional composite dowels

However, the relative thin concrete thickness of thin-covered composite dowels makes it difficult for the pry-out cone to fully expand in the thickness direction, causing the concrete failure to extend to the concrete side surface. This leads to a different failure mode compared to conventional composite dowels. Therefore, a new shear capacity model for thin-covered composite dowels is developed here.

A design model for head studs positioned close to the concrete surface has been proposed in previous studies [40], whose failure mode is similar to that of thin-covered composite dowels, primarily characterized by the peeling of the side concrete cover. Based on extensive experimental data, it can be observed that there is a functional relationship between the shear capacity and the effective height of the head stud. A similar relationship can be inferred for thin-covered composite dowels.

In this study, it is found that the shear capacity of thin-covered composite dowels is not only related to the effective height of the steel dowel but also to the top concrete cover $c_t$. Therefore, the effective embedment depth $h_{ef}$ is used to integrate these two factors, replacing the pry-out cone height $h_{po}$ in conventional composite dowels. Similar to Section 4.4, for the pure shear capacity model of thin-covered composite dowels, the relationship between shear capacity and the size of the pry-out cone in conventional composite dowels is transformed into a relationship with the effective embedment depth, effective distance, concrete thickness, and steel dowel thickness, while retaining the dependence on concrete strength.

Summarizing Sections 5.1–5.4, the complete shear capacity models are proposed in Eq. (28) and Eq. (29) for thin-covered composite dowels with ordinary concrete, and Eq. (30) and Eq. (31) for those with SFRC. The comparison between the FEA results and model predictions is shown in Fig. 27. The results indicate that the model predictions align well with the FEA results, confirming the accuracy of the pure shear capacity model.

$$F_{S,CL} = 1.465 \cdot b_c^{0.705} \cdot t_w^{0.398} \cdot \sqrt{f_{ck}} \cdot L_{e,CL} \cdot h_{ef}^{0.23} \tag{28}$$

$$F_{S,PZ} = 0.278 \cdot b_c^{0.983} \cdot t_w^{0.331} \cdot \sqrt{f_{ck}} \cdot L_{e,PZ} \cdot h_{ef}^{0.26} \tag{29}$$

$$F_{S,CL} = 1.465 \cdot (0.11401 \cdot f_{tk} + 1.59572) \cdot b_c^{0.705} \cdot t_w^{0.398} \cdot \sqrt{f_{ck}} \cdot L_{e,CL} \cdot h_{ef}^{0.23} \tag{30}$$

$$F_{S,PZ} = 0.278 \cdot (0.18519 \cdot f_{tk} + 1.70541) \cdot b_c^{0.983} \cdot t_w^{0.331} \cdot \sqrt{f_{ck}} \cdot L_{e,PZ} \cdot h_{ef}^{0.26} \tag{31}$$

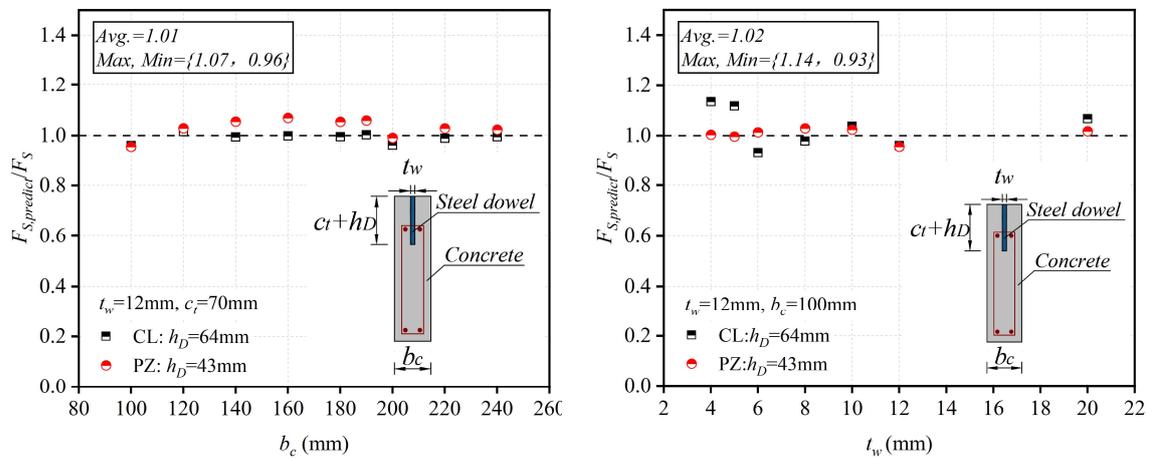

(a) Concrete width (NC)　　　　　(b) Steel thickness (NC)

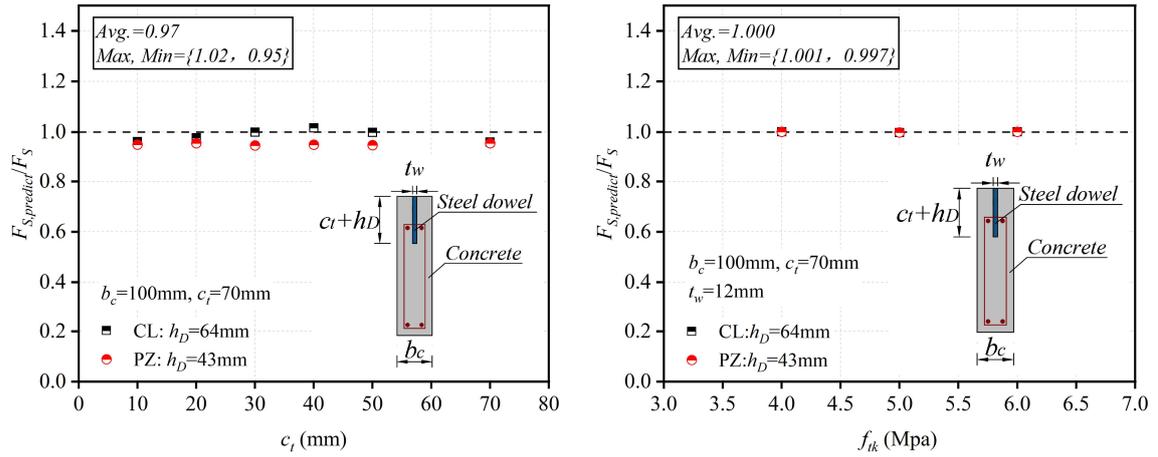

(c) Effective embedment depth (NC)   (d) Tensile strength (SFRC)

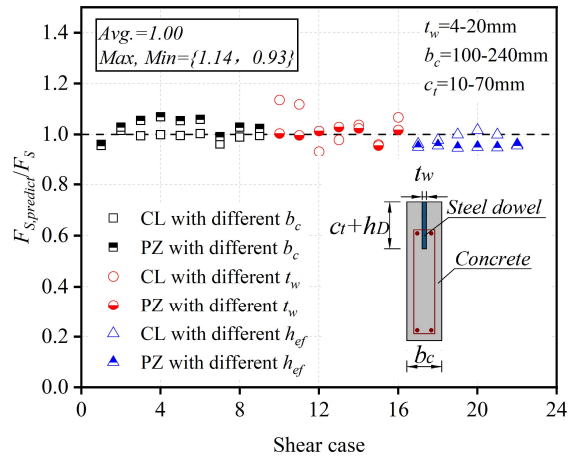

(e) Total

Fig. 27 Comparison between the FEA and the model prediction results

## 6. Tensile-shear coupling capacity of thin-covered composite dowels

In Sections 4.4 and 5.4, the pure tensile and shear capacity models of thin-covered composite dowels were discussed in detail (Eq. (14)-(17) and Eq. (26)-(29)). However, in practical engineering applications, composite dowels are often subjected to both shear and tensile loads simultaneously. For instance, in areas near the negative bending moment regions of bridges, composite dowels serving as shear connectors must withstand tensile-shear coupling loads.

As shown in Fig. 28, under tensile-shear coupling loading, the insufficient thickness of the thin side concrete cover restricts the development of the failure cone in the thickness direction.

Consequently, cracks propagate to the side surface of the concrete, resulting in RCF, which differs from the failure mode in conventional composite dowels.

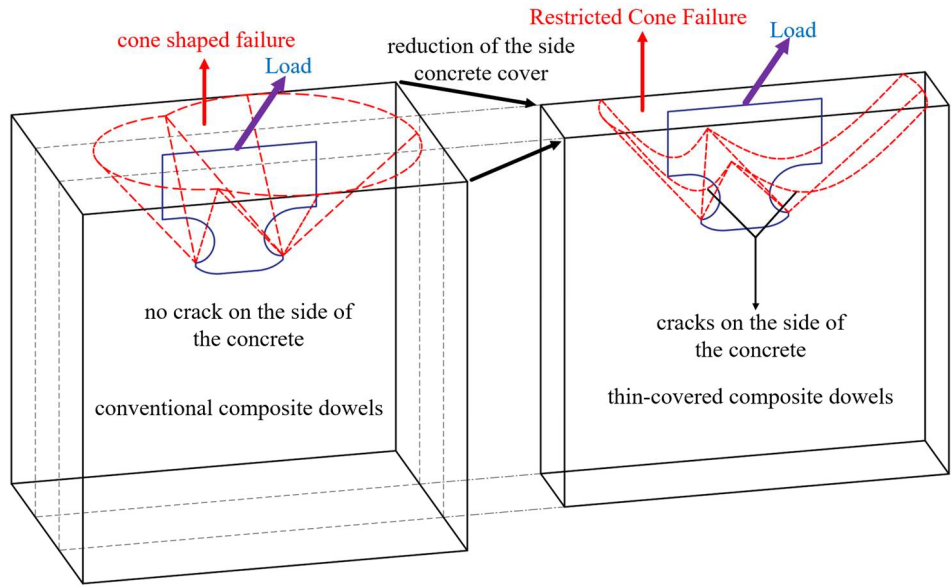

Fig. 28 Comparison of the failure mode between thin-covered composite dowels and conventional composite dowels

Thus, in addition to the pure tensile and shear capacity models, a separate model is necessary to describe the interaction between tensile and shear forces in order to determine the tensile-shear coupling capacity of thin-covered composite dowels. While several mathematical models exist for the interaction of conventional composite dowels, the model presented in Eq. (32) [31] is adopted in this study. In this model, $T$ and $S$ represent the tensile and shear capacities of the thin-covered composite dowels under the tensile-shear coupling load, while $F_T$ and $F_S$ represent the pure tensile and shear capacities, as detailed in Sections 4.4 and 5.4. The coefficient $k$ determines the interaction shape between the two capacities, and its accurate determination is crucial for predicting the tensile-shear coupling capacity.

$$\left(\frac{T}{F_T}\right)^k + \left(\frac{S}{F_S}\right)^k = 1 \qquad (32)$$

Since the key distinguishing feature of thin-covered composite dowels is their thickness dimensions, a parametric analysis is performed with selected representative thickness values to determine $k$. The parameters for this analysis are summarized in Table 7. The concrete thickness $b_c$ ranges from 100-240mm, the steel dowel thickness $t_w$ ranges from 4-20mm, the loading angle $\theta$ varies between 0° and 90°, and the concrete is C50 with a top concrete cover $c_t$ of 70mm, which is considered sufficient.

From Fig. 29, it is observed that the tensile-shear coupling capacity of the composite dowels decreases with the loading angle $\theta$. To calculate $k$, the bearing capacities are normalized as shown in Eq. (33) and Eq. (34), as depicted in Fig. 30. The results show that for CL composite dowels, the value of $k$ ranges from approximately 1.51 to 2.10, while for PZ composite dowels, $k$ ranges from approximately 1.53 to 2.69.

$$S/F_S = \frac{F_{T-S}\cos(\theta)}{F_S} \quad (33)$$

$$T/F_T = \frac{F_{T-S}\sin(\theta)}{F_T} \quad (34)$$

Using the obtained $k$, an engineering empirical model for thin-covered composite dowels is proposed as expressed in Eq. (35). In this model, $T_a$ and $S_a$ represent the actual tensile and shear loads applied to the composite dowel. If the calculated value $M$ from the model is less than 1, the composite dowel is within the safe bearing capacity range. Conversely, if $M$ exceeds 1, the dowel has surpassed its ultimate bearing capacity, indicating that failure is imminent and the dowel requires immediate repair.

$$M = (\frac{T_a}{F_T})^k + (\frac{S_a}{F_S})^k \quad (35)$$

Table 7 Parameter scheme for $b_c$, $t_w$ and $\theta$

| Parameter | Size |
| --- | --- |
| $b_c$/mm | 100, 120, 140, 160, 200, 240 |

| $t_w$/mm | 6, 8, 10, 12, 20 |
| $\theta$/° | 0, 15, 30, 45, 60, 75, 90 |

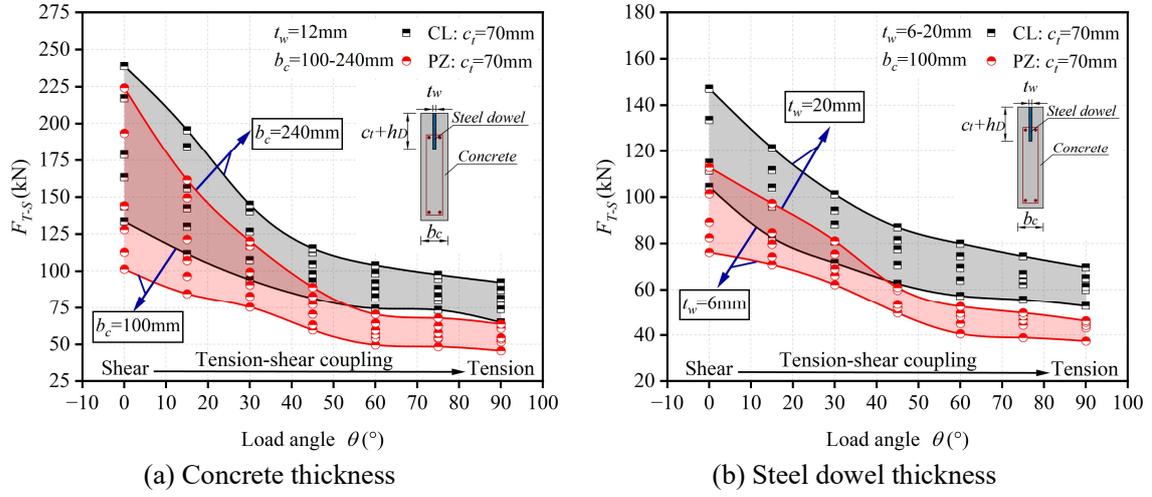

(a) Concrete thickness  (b) Steel dowel thickness

Fig. 29 Tensile-shear coupling capacity of thin-covered composite dowels

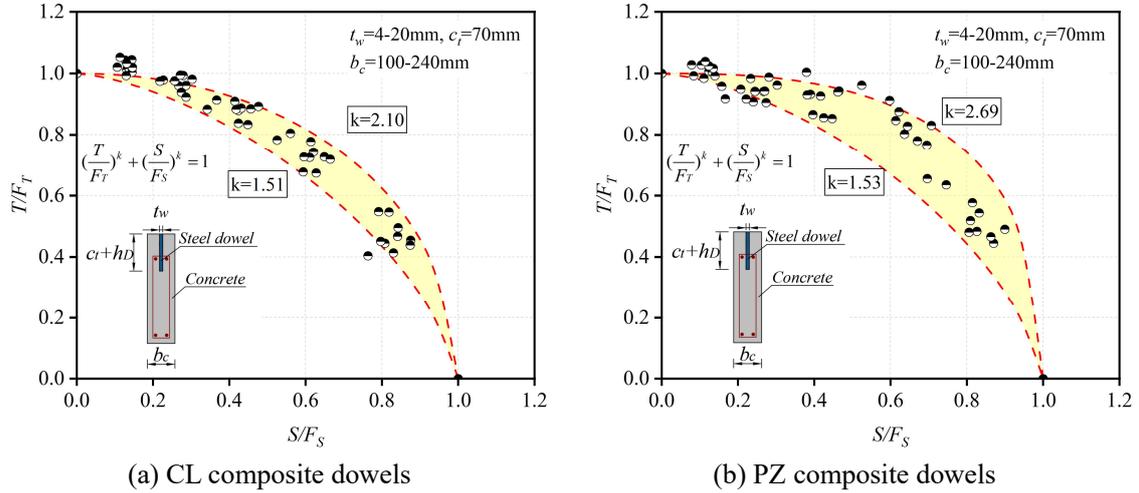

(a) CL composite dowels  (b) PZ composite dowels

Fig. 30 Tensile-shear coupling capacity model of thin-covered composite dowels

## 7. Conclusions

This study provides a comprehensive investigation into the performance of thin-covered composite dowels under pure tensile, shear, and tensile-shear coupling loads. By conducting experiments and parametric studies, the effects of key parameters such as steel dowel thickness, concrete thickness, concrete cover, and the Steel Fiber Reinforced Concrete (SFRC) on dowel bearing capacity have been thoroughly examined. The following conclusions are drawn:

- A series of pull-out tests are designed for thin-covered composite dowels, along with the

development of a novel test rig and special specimens to achieve tensile-shear coupling loading. It is found that the failure mode of thin-covered composite dowels under tensile-shear coupling load differs from conventional composite dowels. Specifically, a *Restricted Cone Failure* (RCF) occurs, and the strengthening effect of SFRC on both the bearing capacity and ductility of thin-covered composite dowels is confirmed.

- The study discusses the difference in failure mechanisms between thin-covered composite dowels and conventional composite dowels. In the case of thin-covered composite dowels, under tensile-shear coupling load, the thin concrete side cover restricts the development of the concrete failure cone in the thickness direction. As a result, cracks extend to the concrete side surface, leading to RCF, a failure mode not observed in conventional composite dowels due to the adequate side concrete cover.

- The tensile, shear, and tensile-shear coupling capacity models for thin-covered composite dowels are developed. The study identifies the key parameter $k$ in the tensile-shear coupling capacity model, with values ranging from 1.51 to 2.10 for CL composite dowels, and from 1.53 to 2.69 for PZ composite dowels. These findings lead to the creation of a theoretical model for the practical design of thin-covered composite dowels.

- The influence of various parameters on the bearing capacity and ductility of thin-covered composite dowels is studied in detail. Among these parameters, concrete thickness and steel dowel thickness emerge as the most important factors. Specifically, for CL composite dowels, increasing concrete thickness from 100 mm to 240 mm improves shear and tensile capacity by 79% and 42%, respectively, while for PZ composite dowels, the improvements are 121% and 40%. Similarly, increasing steel dowel thickness from 4 mm to 20 mm enhances shear and tensile capacity by 101% and 62% for CL composite dowels and by 68% and 32% for PZ composite dowels. However, increasing steel dowel thickness reduces the ductility of thin-covered composite dowels, while increasing concrete thickness improves the ductility of PZ composite dowels but has minimal impact on the ductility of CL composite dowels.

## Acknowledgements

The authors are grateful for the support of China Scholarship Council (Grant No.202206305001).